\documentclass[aps,%twocolumn,
twocolumn,groupedaddress,nofootinbib,nobalancelastpage,nobibnotes]{revtex4}
\pdfoutput=1
\usepackage{amsmath,amsfonts,amssymb,mathrsfs,graphicx,color}
\usepackage[squaren]{SIunits}
\usepackage{verbatim}
\usepackage{enumerate}
\usepackage{graphicx}
\usepackage[hyperfootnotes=false]{hyperref}
\usepackage{xcolor}
\usepackage{slashed}
\usepackage[utf8]{inputenc}
\usepackage[mathscr]{euscript}

\newcommand{\andrea}[1]{{\leavevmode\color{black}#1}}

\newcommand{\andy}[1]{{\leavevmode\color{black}#1}}

\newcommand{\mau}[1]{{\leavevmode\color{black}#1}}

\begin{document}

\title{On the IceCube spectral anomaly}


\author{Andrea Palladino \footnote{andrea.palladino@gssi.infn.it}}
\affiliation{Gran Sasso Science Institute, L'Aquila (AQ), Italy}

\author{Maurizio Spurio \footnote{maurizio.spurio@bo.infn.it}}
\affiliation{Dipartimento di Fisica e Astronomia Universit\`{a} di Bologna and INFN Sezione di Bologna}

\author{Francesco Vissani \footnote{francesco.vissani@lngs.infn.it}}
\affiliation{INFN, Laboratori Nazionali del Gran Sasso, Assergi (AQ) \\ Gran Sasso Science Institute, L'Aquila (AQ), Italy}

%\thankstext[$\star$]{t1}{Thanks to the title}
%\thankstext{e1}{e-mail: andrea.palladino@gssi.infn.it}
%\thankstext{e3}{e-mail: maurizio.spurio@bo.infn.it}
%\thankstext{e2}{e-mail: francesco.vissani@lngs.infn.it}

%\institute{Gran Sasso Science Institute, L'Aquila (AQ), Italy \label{addr1}
   %       \and
    %       Dipartimento di Fisica e Astronomia Universit\`{a} di Bologna and INFN Sezione di Bologna \label{addr3}
    %       \and
       %   INFN, Laboratori Nazionali del Gran Sasso, Assergi (AQ), Italy \label{addr2}
       %          \and
  %        \emph{Present Address:} Street, City, Country\label{addr3}
%}


%\andy{PARTI NUOVE} \\ 
%\del{PARTI CHE VORREI ELIMINARE} \\
%\sposta{PARTI SPOSTATE} \\
%\mau{PARTI RISCRITTE MODIFICATE DI POCO (MAU)} \\

\begin{abstract}
Recently it was noted that different IceCube datasets are not consistent
with the same power law spectrum of the cosmic neutrinos: 
this is the {\em IceCube spectral anomaly}, that suggests that they  observe
a multicomponent spectrum.
In this work, the main possibilities to enhance the description in terms of
a single extragalactic neutrino component are examined. The hypothesis of a
sizable contribution of Galactic high-energy neutrino events distributed as
$E^{-2.7}$ [ApJ 826, 185 (2016)] is critically analyzed and its natural
generalization is considered. The stability of the expectations is studied
by introducing free parameters, motivated by theoretical considerations and
observational facts. 
The upgraded model here examined 
%and called $\nu${\sf gal2} 
has 1)~a Galactic component with different normalization and shape $E^{- 2.4}$; 2)~an
extragalactic neutrino spectrum based on new data; 3)~a non-zero prompt
component of atmospheric neutrinos. The two key predictions of the model
concern the `high-energy starting events' collected from the Southern sky.
%They remain qualitatively unaltered also in the $\nu${\sf gal2} variant: 
The Galactic component produces a softer spectrum and a testable angular
anisotropy. A second, radically different class of models, where the second
component is instead isotropic, plausibly extragalactic and with a
relatively soft spectrum, is disfavored instead by existing observations of
muon neutrinos from the Northern sky and below few 100 TeV.
\end{abstract}

\setcounter{secnumdepth}{4}
\setcounter{tocdepth}{4}

\maketitle

%%%%%%%%%%%%%%%%%%%%%%%%%%%%%%%%%%%%%%%%%%%%%%%%%%%%%%%%%%%%%%%%%%%%%%%%% 
\section{Introduction}
%%%%%%%%%%%%%%%%%%%%%%%%%%%%%%%%%%%%%%%%%%%%%%%%%%%%%%%%%%%%%%%%%%%%%%%%% 

The observations of IceCube \cite{iceini,icesci,ice2pev,icecomb,Passing} provided us with a convincing evidence of cosmic neutrinos. 
The observed events cannot be attributed to cosmic ray that interact with our atmosphere, and this is especially true for the four events with visible energies above the PeV \cite{iceini,ice2pev,Passing}. 
Moreover, the observations are compatible with an isotropic angular distribution of the events, which is a circumstantial clue that the bulk of these cosmic neutrinos has an extragalactic origin.

However, there is no good reason to assume that high energy Galactic neutrinos are entirely absent \cite{viena}. 
A Galactic contribution to the events observed by IceCube has been considered by several authors \cite{nero1,spurio,nero2,troitsk,apj,evoli}. 
%An additional emission of neutrinos from the Galaxy that contributes to %the events observed by IceCube, has been considered by several authors %\cite{nero1,spurio,nero2,troitsk,apj,evoli}. 
%
There are in fact various hints for a Galactic component. E.g., the excess of events in the Southern sky is larger than what is expected assuming isotropy \cite{spurio}. Moreover, assuming a power-law distribution,
\begin{equation}
\frac{d\phi_{\nu_\ell+ \bar\nu_\ell}}{dE}= n\ E^{-\alpha}
\mbox{ with }\ell=e,\mu,\tau
\end{equation}
the high energy starting events (HESE) from the Southern sky require a spectral index $\alpha\approx 2.5$, whereas the passing muons coming from the Northern sky require $\alpha\approx 2.0$  instead \cite{apj}.

The IceCube collaboration attributed to this discrepancy a significance of 3.3$\sigma$ \cite{Passing}, similar to the significance of 3$\sigma$ estimated in \cite{apj}. 
This discrepancy is referred in the following as the IceCube {\em spectral anomaly}.
It strengthens the likelihood that a Galactic neutrino component exists, mainly observable from the Southern hemisphere.

Neutrino oscillations play an important role on cosmic neutrinos, modifying the flavor ratio at Earth with respect to the flavor ratio near the sources. In most production scenario, oscillations imply that the neutrinos of all three flavors have on Earth the same spectra and very similar normalizations \cite{u1,u2,u3,u4,u5,u6,u7,u8}; these considerations are supported by the observations \cite{Palladino:2015zua,w32}.
Unless stated otherwise, in most of this work we focus on the {\em all-flavor} flux--i.e., the sum of the 3 flavors of neutrinos and antineutrinos, that we denote simply as $\phi$.

The hypothesis of a \mau{not negligible} Galactic neutrino emission allows to reconcile the different spectral indexes measured in the two IceCube samples.
The Galactic component adds events only in the Southern sky, being almost absent in the Northern sky \cite{spurio}.
The specific model discussed in \cite{apj} ($\nu${\sf gal1} in the following) corresponds to a  spectrum of the cosmic neutrinos, seen from the Southern sky, given by the sum of two power-laws with fixed spectral indexes, $\alpha=2$ and $\alpha=2.7$, for the  extragalactic and Galactic components, respectively. 

The main aim of this work is to generalize the $\nu${\sf gal1} in a new model, denoted as $\nu${\sf gal2}. It still assumes that the main component of IceCube cosmic neutrinos is of extragalactic origin, with the presence of an important Galactic contribution. 
$\nu${\sf gal2} is motivated by addressing the following important questions:
\begin{enumerate}
\item How crucial is the choice of the shape of the Galactic component? 
What happens if it is closer to the one indicated by observations of $\gamma$-rays and some models of cosmic ray propagation, i.e., $\alpha\approx 2.4-2.6$?
\item What happens changing the normalization of the Galactic component?  
\item What is the effect of prompt neutrinos? (This was neglected in \cite{apj})
\andy{\item A model with two extragalactic components is also compatible with observations?}
\end{enumerate} 
We derived the $\nu${\sf gal2} model after exploring these questions, quantifying the expectations and considering constraints, and assessing the stability of its predictions.

In the Sect.~\ref{sec:unk} we begin by examining the experimental dataset provided by the IceCube collaboration.
Then in Sect.~\ref{high} we will discuss the theoretical prediction for the spectra of different populations of neutrinos. 
The model $\nu${\sf gal2} is presented and discussed in Sect.~\ref{pv2sum}.
An alternative model, that includes two extragalactic components instead, is discussed in Sect.~\ref{cacaton}. 
We draw our conclusions in Sect.~\ref{conc}.

%%%%%%%%%%%%%%%%%%%%%%%%%%%%%%%%%%%%%%%%%%%%%%%%%%%%%%%%%%%%%%%%%%%%%%%%% 
\section{Experimental observations}\label{sec:unk}
%%%%%%%%%%%%%%%%%%%%%%%%%%%%%%%%%%%%%%%%%%%%%%%%%%%%%%%%%%%%%%%%%%%%%%%%% 

{There are two dataset available from the IceCube experiment \mau{in which neutrinos of astrophysical origin have been observed. 
The first one refers to upward going muons induced by charged current (CC) interaction of $\nu_\mu+\overline \nu_\mu$ crossing the Earth.}
% to the classic way to detect neutrinos, i.e. observing the muon 
% produced by their interaction crossing the Earth. 
These neutrinos \mau{arriving in IceCube are originating} from the Northern hemisphere. 
\mau{In the following, we refer to this sample} with the name of "Passing muon". 
The second dataset is characterized by neutrinos that interact into \mau{a fiducial volume of the detector, with a contained interaction vertex}; we refer to this sample as "High Energy Starting Event" (HESE).}

%%%%%%%%%%%%%%%%%%%%%%%%%%%%%%%%%%%%%%%%%%%%%%%%%%%%%
\subsection{Passing muons and extragalactic spectrum}
%%%%%%%%%%%%%%%%%%%%%%%%%%%%%%%%%%%%%%%%%%%%%%%%%%%%%

IceCube used data from 2009 through 2015 to measure CC upgoing muon neutrino events, with the field of view restricted to the Northern hemisphere \cite{Passing}.
The highest energy sample (with reconstructed energy above 190 TeV) corresponds to 29 events that exclude a purely atmospheric origin at 5.6$\sigma$ significance. 
The corresponding cosmic muon flavor (neutrino+antineutrino) flux, estimated from these data, was obtained with a fit to the power-law:
\begin{equation}
\frac{d\phi_{\nu_\mu+\bar\nu_\mu}}{dE}=\frac{0.90^{+0.30}_{-0.27} \times 10^{-18} }{\rm GeV \ cm^{2} \ s \ sr } \left(\frac{E}{\rm 100 \ TeV} \right)^{-2.13 \pm 0.13}  
\label{piril}
\end{equation}
%This slope is consistent with the theoretical expectations discussed %above. 
No correlation with known $\gamma$-ray sources was found by analyzing the arrival directions of these 29 events.

%%%%%%%%%%%%%%%%%%%%%%%%%%%%%%%%%%%%%%%%%
\subsection{High Energy Starting Events\label{pipr}}
%%%%%%%%%%%%%%%%%%%%%%%%%%%%%%%%%%%%%%%%%
%An important dataset collected by IceCube concerns the contained events, better known as high energy starting events (HESE). 
In 4 years of data taking 54 HESE have been detected, that are classified as 14 tracks and 39 showers  events (1 of them is not identified). Among them, 16 events come from the Northern hemisphere, 37 events come from the Southern hemisphere and 1 event was detected with declination equal to zero. These events have a deposited energy above 30 TeV and the most energetic HESE deposited an energy of 2 PeV into the detector. 
%Also other 2 events with deposited energy above 1 PeV have been detected.  

A fraction of these events is due to background, in particular to atmospheric neutrinos produced in $\pi$ and $K$ decay. The background spectrum is approximatively described as a power-law with spectral index $\alpha=3.7$ and normalization known with an uncertainty of about 25\% \cite{icecomb}. These background neutrinos come both from Southern and Northern hemisphere. 
In addition to, there are also atmospheric muons produced especially in charged mesons decay.
% with the same spectral shape. NON INTERESSA
In IceCube, if correctly reconstructed, atmospheric muons come only from Southern hemisphere, because muons produced in the other hemisphere are absorbed by the Earth. 

The remaining part of events is attributed to astrophysical neutrinos and it is described by means of an isotropic distribution, whose all-flavor energy spectrum is a power-law with,
\begin{equation}
\frac{d\phi}{dE}= \frac{6.7^{+1.1}_{-1.2} \times 10^{-18}}{\rm GeV \ cm^{2} \ s \ sr } \left(\frac{E}{\rm 100 \ TeV} \right)^{-2.50 \pm 0.09}  
\label{trompon}
\end{equation}

Although the bulk of HESE events seen  from the Southern sky suggest a power-law spectrum with spectral index  $\alpha\approx 2.5$, the subset of high energy events
is in agreement with a much harder spectrum and more precisely, with the same distribution suggested by the passing muons: 
See Fig.~6 of \cite{apj}
and Fig.~5 of \cite{Passing}, and discussions therein.
In other words,  
%\begin{quote}
the flux of the highest energy HESE events observed from the Southern sky is compatible with a hard $\alpha\approx 2$ spectrum.
%\end{quote} 
 
\subsection{Issues}
%%%%%%%%%%%%%%%%%%%%%
Due to the approximate isotropy of HESE, they are usually believed to be extragalactic neutrinos. 
These assumptions, however, lead immediately to various problems \cite{spurio,apj}, and noticeably, 
1) the corresponding extragalactic 
$\gamma$-ray flux exceeds the 
limits sets by the observations \cite{apj};
2) the spectrum of Eq.~\ref{trompon} 
is different from the flux obtained by the passing muons; 
3) conversely, the flux of Eq.~\ref{piril} is not able  to reproduce the number of events observed in the Southern hemisphere. 

The model $\nu${\sf gal2}, described in Sect.~\ref{pv2sum}, provides a solution to these issues.

%%%%%%%%%%%%%%%%%%%%%%%%%%%%%%%%%%%%%%%%%%%%%%%%%%%%%%%%%%%%%%%%%%%%%%%%% 
\section{Considerations on the high energy neutrino spectrum}
\label{high} 
%%%%%%%%%%%%%%%%%%%%%%%%%%%%%%%%%%%%%%%%%%%%%%%%%%%%%%%%%%%%%%%%%%%%%%%%% 

%%%%%%%%%%%%%%%%%%%%%%%%%%%%%%%%%%%%%%%%%%%%%%%%%%%%%%%%%%
%\subsection{Expectations on the extragalactic neutrinos}
\subsection{The spectral index of extragalactic neutrinos}
%%%%%%%%%%%%%%%%%%%%%%%%%%%%%%%%%%%%%%%%%%%%%%%%%%%%%%%%%%

\mau{ An hard energy spectrum $\approx E^{-\alpha}$  for extragalactic neutrinos, with $\alpha\approx 2.0$, is motivated by models of cosmic rays production at sources in the framework of the Fermi acceleration mechanism.}
A very well-known case is the generic model developed in Waxman \& Bahcall~\cite{wb}, where the neutrino spectrum is assumed to have $\alpha=2.0$. Note that this assumption was used in the first IceCube fits~\cite{icesci}.

Similar results have been obtained in other specific models. 
E.g., the shape of the extragalactic component in the starburst galaxies model of \cite{lw} is 
\begin{equation}\frac{d\phi_{\mbox{\tiny sfg}}}{dE}\propto E^{-2.15\pm 0.1}\end{equation}
and this extends till  0.3 PeV at least. 
A generic bound, $\alpha<2.1-2.2$ was derived by the authors of \cite{Murase:2013rfa}, assuming hadro-nuclear (pp) scenarios for neutrino production. 
Also in very different models, as \cite{pado2,righi}, based on the hypothesis that the neutrino sources are (some types of) BL Lac/blazars, the power spectrum of the extragalactic neutrino component has a very hard spectral index, close to $\alpha\approx 2$.
%\footnote{
The latter kind of models is  apparently motivated by the {\em Fermi}-LAT  measurements obtained at much lower energies. See \cite{winter} for a principled discussion of the expectations from $p\gamma$ sources, in which different target photon distributions are considered.

According to the previous theoretical and experimental arguments, the spectral index of the extragalactic component is reasonably expected in the conservative (wide) range
\begin{equation}
\alpha_{\mbox{\tiny eg}}\in [1.9\ ,\ 2.3]
\end{equation} 
 
The information arising from theoretical models on the flux normalization are more uncertain; a stronger constraint arises from the IceCube data itself.

%%%%%%%%%%%%%%%%%%%%%%%%%%%%%%%%%%%%%%%%%%%%%%%%%%%
\subsection{Expectation on the Galactic component}
%%%%%%%%%%%%%%%%%%%%%%%%%%%%%%%%%%%%%%%%%%%%%%%%%%%

Here we discuss the core of the $\nu${\sf gal2} model, namely the component attributed to Galactic neutrinos (defined by its spectral index and its normalization). 

\subsubsection{The spectral index of the Galactic component}\label{slopa}
%%%%%%%%%%%%%%%%%%%%%%%%%%%%%%%%%%%%%%%%%%%%%%%%%%%%%%%%%%%%%%%%%
Cosmic rays till few PeV impinge the Earth with a power-law distribution with $\alpha\approx 2.7$. This is thought to be the result of a harder injection spectrum, modified after the propagation inside the Galaxy. 
However, neutrinos are plausibly produced in collisions close to their cosmic sources, and in this case, they should reflect the injection spectrum. 
According to this physical picture, neutrinos should have a spectrum harder (i.e., smaller) than $\alpha\approx 2.7$. 

There are several attempts to estimate the Galactic neutrino flux. Two theoretical estimations~\cite{candia,guo}  find an interesting level of the diffuse neutrino flux from the Galactic plane: in both cases the  power-law approximation of this flux,  $\phi\propto E^{-\alpha}$ has a spectral index $\alpha=2.5-2.6$ in the region from some 10 of TeV till several 100 of TeV.

A phenomenological model characterized by radially dependent cosmic-ray transport properties has been recently used to compute the $\gamma$-ray and neutrino diffuse emission of the Galaxy \cite{gagge}.
The model, designed to reproduce both Fermi-LAT $\gamma$-ray data and local cosmic ray observables, naturally reproduces the anomalous TeV diffuse emission observed by Milagro in the inner Galactic plane. 
Above 100 TeV it predicts a neutrino flux with a spectral index $\alpha \approx 2.4\div 2.5$ that is about two to five times larger than the neutrino flux obtained with conventional models in the Galactic Center region. This neutrino flux explains up to 25\% the flux measured by IceCube.

%%%%%%%%%%%%%%%%%%%%%%%%%%%%%%%%%%%%%%%%%%%%%%%%%%%%%%%%%%%%%%%%%%
\begin{figure}[t]
\centering
\includegraphics[width=0.8\linewidth]{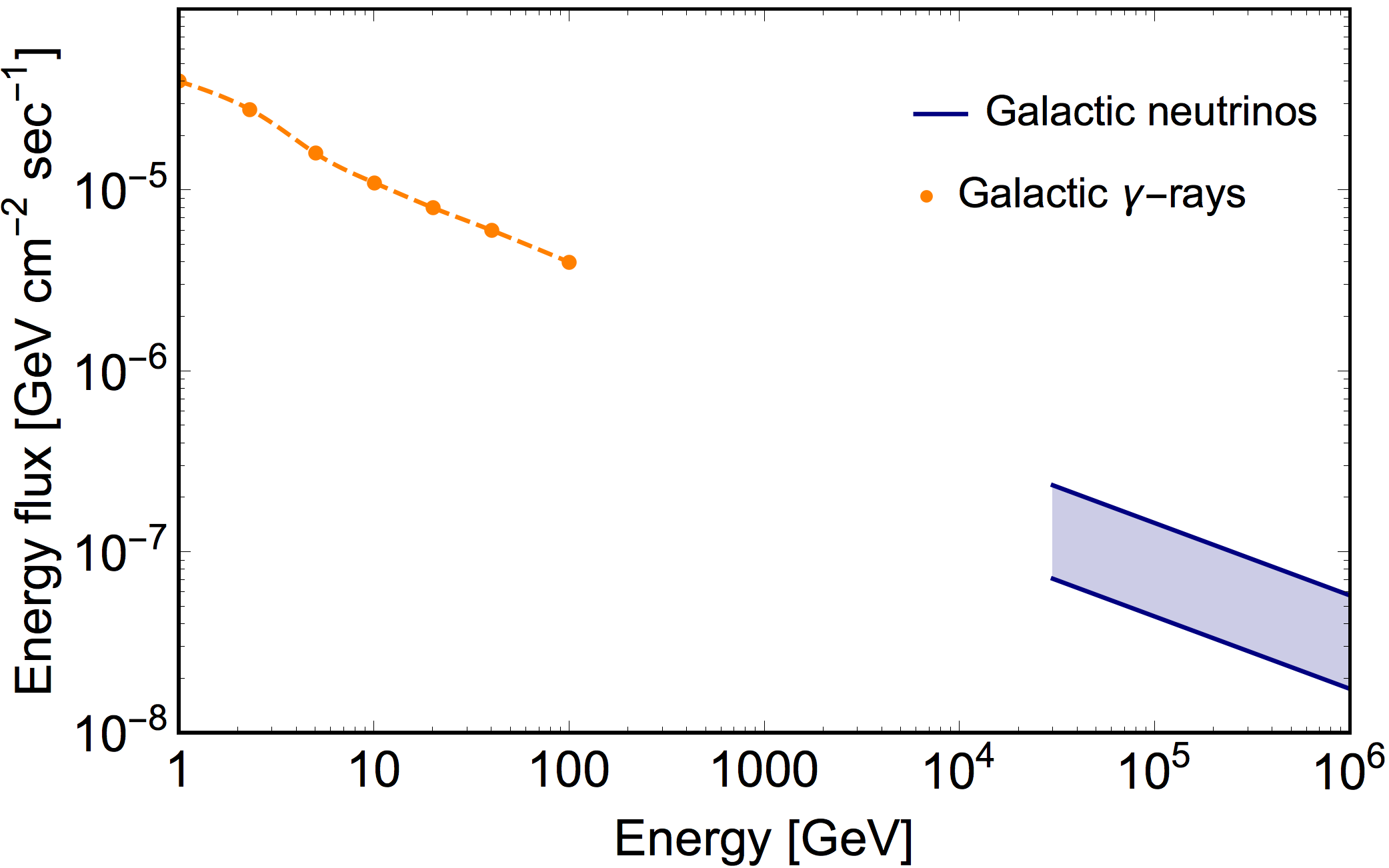}
\caption{\textit{In orange, the spectrum of diffuse gamma-ray emission from the inner Galaxy ($-80^\circ\le l \le 80^\circ, -8^\circ\le b \le 8^\circ$),  measured by {\em Fermi}-LAT. In blue, the all-flavor Galactic neutrino component of the $\nu${\sf gal2} model with its uncertainties.}}
\label{fig:fff}
\end{figure}
%%%%%%%%%%%%%%%%%%%%%%%%%%%%%%%%%%%%%%%%%%%%%%%%%%%%%%%%%%%%%%%%%%

Moreover, the spectral index of the $\gamma$-ray emission, observed at lower energies from the Milky Way, resembles $\alpha\approx 2.4-2.6$ more than $\alpha\approx 2.7$. This can be seen from Fig.~\ref{fig:fff}, where we compare a neutrino spectrum with spectral index $\alpha=2.4$ with the gamma rays seen by {\em Fermi}-LAT; 
see~\cite{apj} for references and compare with Fig.~8 therein. 
%\andyb{NON SO SE E' IL CASO DI MOSTRARE QUESTA FIGURA QUI. NON ABBIAMO ANCORA DEFINITO IL FLUSSO DI NEUTRINI GALATTICI}

According to the above arguments, the spectral index of the Galactic component can be assumed to belong to the range,
\begin{equation}
\alpha_{\mbox{\tiny g}}\in [2.4\ ,\ 2.7]
\end{equation}
that does not overlap with the corresponding range for the extragalactic neutrinos. Moreover, in order to maximize the difference with the assumptions of $\nu${\sf gal1} model, but keeping in mind the above theoretical and observational indications,  
we will assume the extremal value $\alpha_{\mbox{\tiny g}}=2.4$.

\subsubsection{The normalization of the Galactic component}\label{norma}
%%%%%%%%%%%%%%%%%%%%%%%%%%%%%%%%%%%%%%%%%%%%%%%%%%%%%%%%%%%%%%%%%%%%%%%%%
%The key assumption is the existence of a significant flux of Galactic %neutrinos. 
The normalization factor for the flux of Galactic neutrinos is poorly constrained by theoretical models, although experimental indications of a non-null value exist.
%not reliably known 
%theoretically,  but there are good reasons to believe that this %component exists. 
A part the mentioned IceCube spectral anomaly, the first hints arises from the intense flux of $\gamma$-rays observed by {\em Fermi}-LAT and Agile from the Galactic plane and its surroundings.
Assuming that a fraction of the signal is due to hadronic mechanisms, a comparable flux of high-energy neutrinos is foreseen. 
Second, the angular distribution of HESE is compatible with the isotropy, but the assumption of a certain degree of anisotropy near to the Galactic plane improves the agreement with the observations \cite{troitsk,apj}.

\begin{figure}[b]
\centering
\includegraphics[scale=0.3]{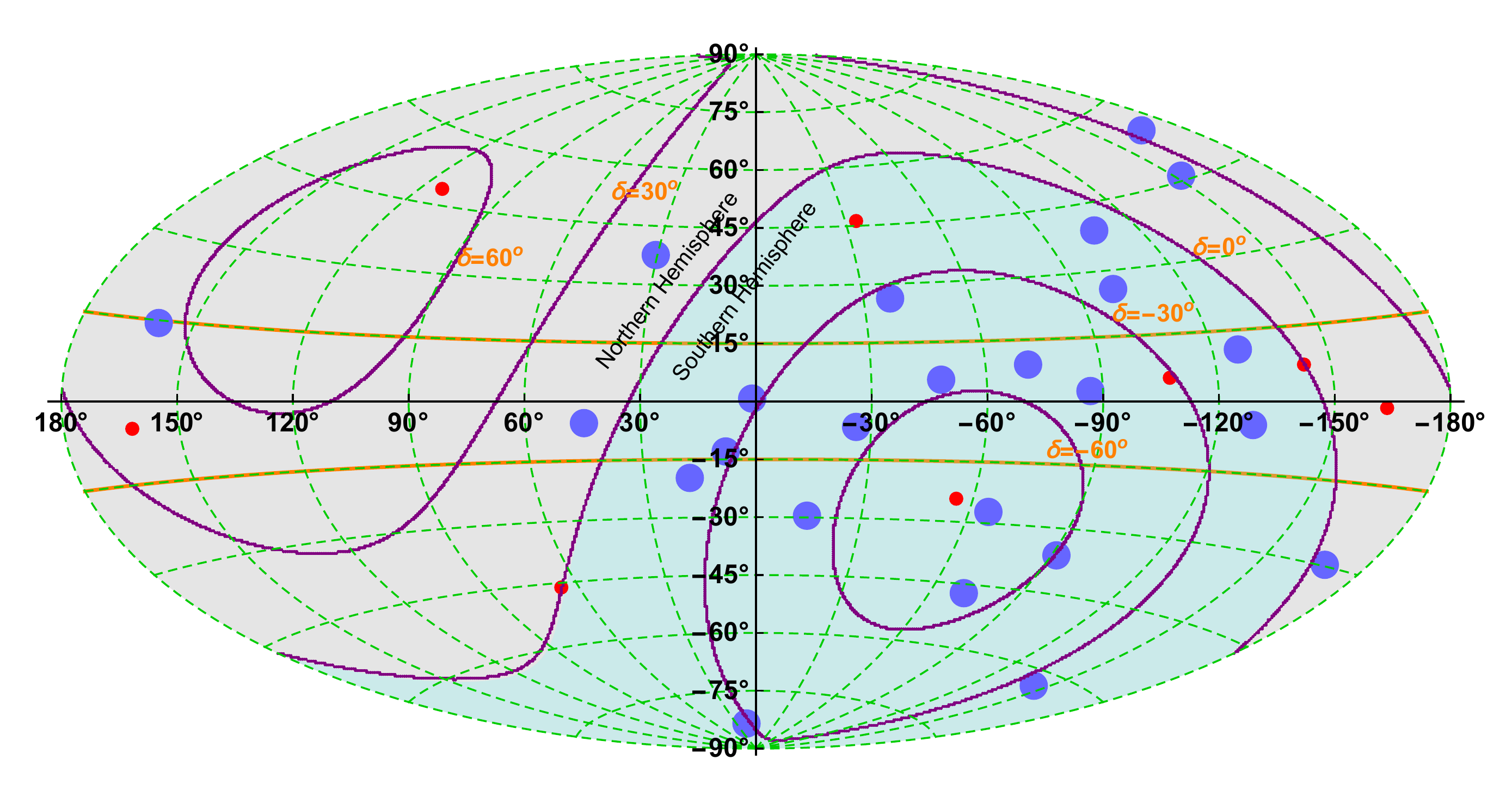}
\caption{\textit{\mau{Distribution of the position of HESE (blue: shower events; red: tracks)} with deposited energy $> 60$ TeV represented in Galactic coordinates. 
\mau{The angular uncertainties are not represented; the showers have much larger angular uncertainty with respect to the tracks.} }}
\label{hesegal}
\end{figure}

Fig.\ref{hesegal} shows the position on the sky (Galactic coordinates) of HESE.  To reduce the atmospheric background, only events with deposited energy $> 60$ TeV are used. 
The angular precision of shower events is $\sim 15^\circ$. Events with $|b| \leq 15^\circ$ are compatible to be originated in the Galactic plane. 

Referring to Fig.\ref{hesegal}, there are 18 events in the region $|b| >15^\circ$. The solid angle covered by this region is approximatively three times larger than the corresponding region with $|b| \le 15^\circ$. Thus, roughly six events are expected in the low longitude region assuming an isotropic detector response. 
The 14 observed events represents a $\sim 2\sigma$ excess with respect to this hypothesis. 

%%%%%%%%%%%%%%%%%%%%%%%%%%%%%%%%%%%%%%%%%%%%%%%%%%%%%%%%%%%%%%%%%%
%%%%%%%%%%%%%%%%%%%%%%%%%%%%%%%%%%%%%%%%%%%%%%%%%%%%%%%%%%%%%%%%%%

These experimental hints supporting the existence of a Galactic component are discussed in different ways in several works \cite{spurio,nero2,apj,candia,evoli}.
E.g., the paper~\cite{evoli} finds that it is theoretically possible to have similar signals in IceCube. Their model called {\em Case C} predicts 1 event per year, using assumptions on the diffusion coefficient of cosmic rays. 

Thus, we will explore the hypothesis that the Galactic neutrino emission yields an observable signal, namely, few HESE events per year in IceCube. As for the case of the extragalactic flux, we extrapolate the normalization of the Galactic component from the data itself.

\subsubsection{The spatial extension of the Galactic component}
%%%%%%%%%%%%%%%%%%%%%%%%%%%%%%%%%%%%%%%%%%%%%%%%%%%%%%%%%%%%%%%
The Galactic flux was considered isotropic and seen only from the Southern hemisphere in \cite{apj}. 
This is a minimal hypothesis, useful to evaluate the total number of HESE from Galactic origin.

A more detailed scenario considers also the spatial extension of the Galactic component, taking into account observational constraints derived from the ANTARES \cite{ANTAnim} neutrino telescope.
In particular, as discussed in Sec. \ref{res}, the Galactic neutrino flux is assumed to be isotropic within a region (in Galactic coordinates) of latitude $|b|\le b^*$ and longitude $|\ell|\le \ell^*$.

%%%%%%%%%%%%%%%%%%%%%%%%%%%%%%%%%%%%%%%%%%%%%%%%%%%
\subsection{Prompt neutrinos}\label{paralla}
%%%%%%%%%%%%%%%%%%%%%%%%%%%%%%%%%%%%%%%%%%%%%%%%%%%
Even if a component in the atmospheric neutrino originating from the prompt decay of very heavy mesons (typically containing a charmed quark) is expected from standard cosmic ray interactions with atmospheric nuclei, this flux of \textit{prompt neutrinos} is not yet measured.

Prompt neutrinos are expected with an energy spectrum also approximated by a power-law, with spectral index $\alpha\approx 2.7$. 
The flavor composition \mau{of prompt neutrinos in a neutrino telescope} is $\nu_e:\nu_\mu:\nu_\tau \approx 1:1:0$ and it is different from what is expected for cosmic neutrinos produced in pion decay, that have a flavor composition at Earth of $\nu_e:\nu_\mu:\nu_\tau \approx 1:1:1$. 
The experimental difficulty to identify tau neutrino interactions and the uncertainties on the spectral index of cosmic neutrinos reduce the possibility to distinguish among cosmic and prompt neutrinos using the information on the flavor of detected events, as discussed in \cite{u8} and \cite{Palladino:2015zua}. 

IceCube has recently investigated the existence of a prompt component in the measured $\nu_\mu$ \cite{Passing} in the region of energy below few 100 of TeV. From the no evidence, 90\% C.L. upper limit on the normalization predicted by the reference calculation \cite{Enberg:2008te} (denoted as ERS) is set to $0.50\times$ ERS.
More recent calculations predict smaller fluxes \cite{p1,p2,p3}, that are not yet significantly probed instead. 
 
Differently from the $\nu${\sf gal1} model\cite{apj}, we assume here the existence of a prompt neutrino component in addition to cosmic neutrinos.
In order to maximize the effect of the contribution from prompt neutrinos without exceeding severely the constraints from theory and observations, we will assume as normalization term the 90\% c.l. upper limit $0.50\times$ ERS set by IceCube.

%%%%%%%%%%%%%%%%%%%%%%%%%%%%%%%%%%%%%%%%%%%%%%%%%%%%%%%%%%%%%%%%%%%%%%%%% 
\section{$\nu${\sf gal2} model}\label{pv2sum}
%%%%%%%%%%%%%%%%%%%%%%%%%%%%%%%%%%%%%%%%%%%%%%%%%%%%%%%%%%%%%%%%%%%%%%%%% 
\subsection{The recipes} \label{sommariomod}

Here, we summarize the result of the discussion of the previous sections to define the $\nu${\sf gal2} model.
Besides the conventional atmospheric neutrinos, we consider three other components: 
the isotropic extragalactic component, the diffuse Galactic component (presumably enhanced near the Galactic plane) and the prompt (atmospheric) neutrinos. In addition:
\begin{enumerate}
\item We use the IceCube measurement of the flux of passing muons to describe the extragalactic component $\phi_{\mbox{\tiny eg}}$.
The flux is assumed to be isotropic, with a normalization known within 30\% and the spectral index within $\approx 6\%$; 

We assume that the total extragalactic neutrino flux is given by the spectrum reported in Eq.~\ref{piril} multiplied by \mau{a factor of three}, in order to account for the three neutrino flavors. At the highest energies, this component is the dominant one.

\item {We assume the existence of a diffuse Galactic component $\phi_{\mbox{\tiny g}}$}, \mau{mainly originating} from the Southern hemisphere.
%\del{due to the position of the Earth into the Galaxy \cite{apj}.
%We expect that this kind of neutrinos are close to the Galactic plane. }
\mau{This Galactic component} produces the observed asymmetry between HESE South and HESE North, \mau{and between HESE and} the passing muons, softening the spectrum of Southern sky below $\approx$ 100 TeV. 
We assume (Sect.~\ref{slopa}) that its spectral index is $\alpha_{\mbox{\tiny g}}=2.4$.  

We estimate the normalization factor for the Galactic component considering the {\em angular distribution} of the HESE events arising from the Southern sky, and more precisely, the distribution in the Galactic latitude (see of \cite{apj} and Sect.~\ref{norma}).
From that analysis we found that a fraction $0.26 \pm 0.15$ of the 23 HESE observed from the Southern hemisphere, with deposited energy above 60 TeV, can be of Galactic origin. It follows that our normalization is chosen to reproduce the expected number of Galactic HESE, i.e. 
\begin{equation}
N_{\mbox{\tiny g}}=6.0 \pm 3.5\mbox{ in 4 years} 
\end{equation}

This assumption is compatible within errors with that used for $\nu${\sf gal1} model and it not exceeds the diffuse flux limit derived from the ANTARES neutrino telescope from neutrinos originating in the Southern sky \cite{ANTAdif}. In addition, this does not contradict the theoretical expectation of 1 event per year \cite{evoli} within 1$\sigma$ and even less in view of theoretical  uncertainties. 

\item We assume the presence of a prompt atmospheric neutrino component $\phi_{\mbox{\tiny prompt}}$. This is isotropic and with a power-law energy spectrum, with spectral index $\alpha=2.7$ and with 
normalization equal to the 90\% c.l. IceCube upper limit, $0.50\times$ ERS. 
\end{enumerate}

 Following these considerations, the all-flavor flux 
 in the $\nu${\sf gal2} model is given by,
\begin{equation}
\frac{d\phi}{dE}=\sum_i  \frac{F_{0,i} \times 10^{-18}}{\rm GeV \ cm^{2} \ s \ sr } \left(\frac{E}{\rm 100 \ TeV} \right)^{-\alpha_i}  
\end{equation}
where the coefficients of normalizations and the spectral indexes are,
\begin{equation} %\label{eq:norma}
\begin{array}{llll} \label{eq:PV2}
%\phi_{0,\mbox{\tiny conv}} &= 2.4 \pm  0.6 & & \alpha_{\mbox{\tiny conv}} = -3.7  \\
F_{0,\mbox{\tiny eg}} &= 3 \times 0.90^{+0.30}_{-0.27} & &\alpha_{\mbox{\tiny eg}} = 2.13 \pm 0.13  \\
F_{0,\mbox{\tiny prompt}} &= 0.6 \pm 0.3 && \alpha_{\mbox{\tiny prompt}} = 2.7  \\
F_{0,\mbox{\tiny g}} &= 1.5 \pm 0.8 && \alpha_{\mbox{\tiny g}} = 2.4 
\end{array} 
\end{equation}

%%%%%%%%%%%%%%%%%%%%%%%%%%%%%%%%%%%%%%%%%%%%%%%%%%%%%%%%%%%%%%%%%%
\begin{table}[t]
\caption{\textit{Summary of the number of HESE events expected for each component in four years. The $\mu$ background and the conventional background are taken from \cite{ice2pev} and scaled with the exposure time. The other predictions are obtained using the fluxes \mau{derived here in Eq. \ref{eq:PV2}} ($\nu${\sf gal2} model) and the published effective areas for HESE.}   }
\begin{center}
\begin{tabular}{cccc}
\hline
\bf{Component} & \bf{North} & \bf{South} &\bf{Sum} \\
\hline
Extragalactic \cite{Passing} & $8.8 \pm 1.7 $ & $16.1^{+1.5}_{-1.9}$ & $24.9^{+3.2}_{-3.6}$  \\
Galactic \cite{apj}& $\approx 0$ & $6.0 \pm 3.5$  & $6.0 \pm 3.5$  \\
Prompt \cite{Passing,Enberg:2008te}& $1.5 \pm 0.8$&  $2 \pm 1$ & $3.5 \pm 1.8$  \\
Atmospheric $\mu$ \cite{ice2pev} & 0 & $12.4 \pm 6.2 $& $12.4 \pm 6.2$ \\
Conventional $\pi$/K \cite{ice2pev} & $6.2 \pm 1.9 $& $3.6 \pm 1.2$ & $9.8 \pm 3.1 $  \\
\hline
Total &$16.5 \pm 2.7 $ & $40.1 \pm 7.5$ &$56.6 \pm 8.7$ \\
Observed \cite{ice3} &16.5 & 37.5  & 54 \\
\hline
\end{tabular}
\end{center}
\label{tabev}
\end{table}% 

\begin{table*}[t]
\andy{\caption{Fraction of Galactic flux, as a function of its extension in longitude $\ell$, that can be seen in different intervals of declination. The hypothesis is that the flux is isotropic in the region of Galactic latitude $|b| \leq 4^\circ$ and Galactic longitude $|\ell| \leq 30^\circ,50^\circ \mbox{or } 70^\circ$.}}
\begin{center}
\begin{tabular}{ccccccc}
\hline
Extension $\delta$ & $[60,90)$ & $[30,60)$& $[0,30)$& $[-30,0)$& $[-60,-30)$ & $[-90,-60)$\\
\hline
$|b|^*=4^\circ , |\ell|^*=30^\circ$ & 0\% & 0\% & 0\% & 52\% & 48\% & 0\% \\
$|b|^*=4^\circ , |\ell|^*=50^\circ$  & 0\% & 0\% & 17\% & 34\% & 43\% & 6\% \\
$|b|^*=4^\circ , |\ell|^*=70^\circ$  & 0\% & 2\% & 25\% & 24\% & 34\%& 15\% \\
\hline
\end{tabular}
\end{center}
\label{GCx}
\end{table*}

%%%%%%%%%%%%%%%%%%%%%%%%%%%%%%%%%%%%%%%%%%%%%%%%
\subsection{Results}\label{res}
%%%%%%%%%%%%%%%%%%%%%%%%%%%%%%%%%%%%%%%%%%%%%%%%

Using the fluxes and the total HESE effective areas \cite{ice3}, we calculate the total number of expected events, with the standard formula written symbolically, 
\begin{equation}
N= 4 \pi \ T \int_0^\infty \! dE\ \sum_{\ell}
\frac{d\phi_{\nu_\ell+\bar\nu_\ell}}{dE}(E)\ A_{\mbox{\tiny eff},\ell}^{\mbox{\tiny HESE}}(E)
\label{rate}
\end{equation}
where we sum over the flavors $\ell=\mbox{e},\mu,\tau$. 
The cosmic components are equipartitioned in the three flavors while the prompt neutrinos are equipartitioned among electron and muon neutrinos.
%\del{We will be interested to distinguish between the events measured in the South and North sky;}
The effective areas for events arising from the South and North sky differ, in particular because of the absorption of very high-energy neutrino in the Earth. 
When the events are separated for each hemisphere, the solid angle factor in Eq. \ref{rate} is $2\pi$.  
The time exposure for HESE data is assumed to be $T=4$ yr.

In Table~\ref{tabev} the contributions from the different components are reported. 
In the case of prompt and Galactic neutrinos, the uncertainty is simply given by the uncertainty on the flux normalization. 
The Galactic component is obtained using the HESE effective areas from the Southern sky.
For extragalactic neutrinos the uncertainty is obtained evaluating the expected number of events obtained at best fit and with the highest (lowest) normalization and the highest (lowest) shape. 
This procedure is justified by the strong correlation between the normalization and the spectral index, that can be noticed in the Fig.6 of \cite{Passing}. 
The expected background \mau{(atmospheric $\mu$; conventional neutrinos from charged $\pi, K$ decays)} is estimated using the \mau{values reported in} \cite{ice2pev} and scaled with the considered exposure time $T$. The uncertainty $\delta$ on the total number of events is obtained combining the various uncertainties $\delta_i$ in quadrature, $\delta=\sqrt{\sum_i \delta_i^2}$.

%%%%%%%%%%%%%%%%%%%%%%%%%%%%%%%%%%%%%%%%%%%%%%%%%%%%%%%%%%%%%%%%%%%%%%%
\begin{figure*}[t]
\centering
\includegraphics[width=0.495\linewidth]{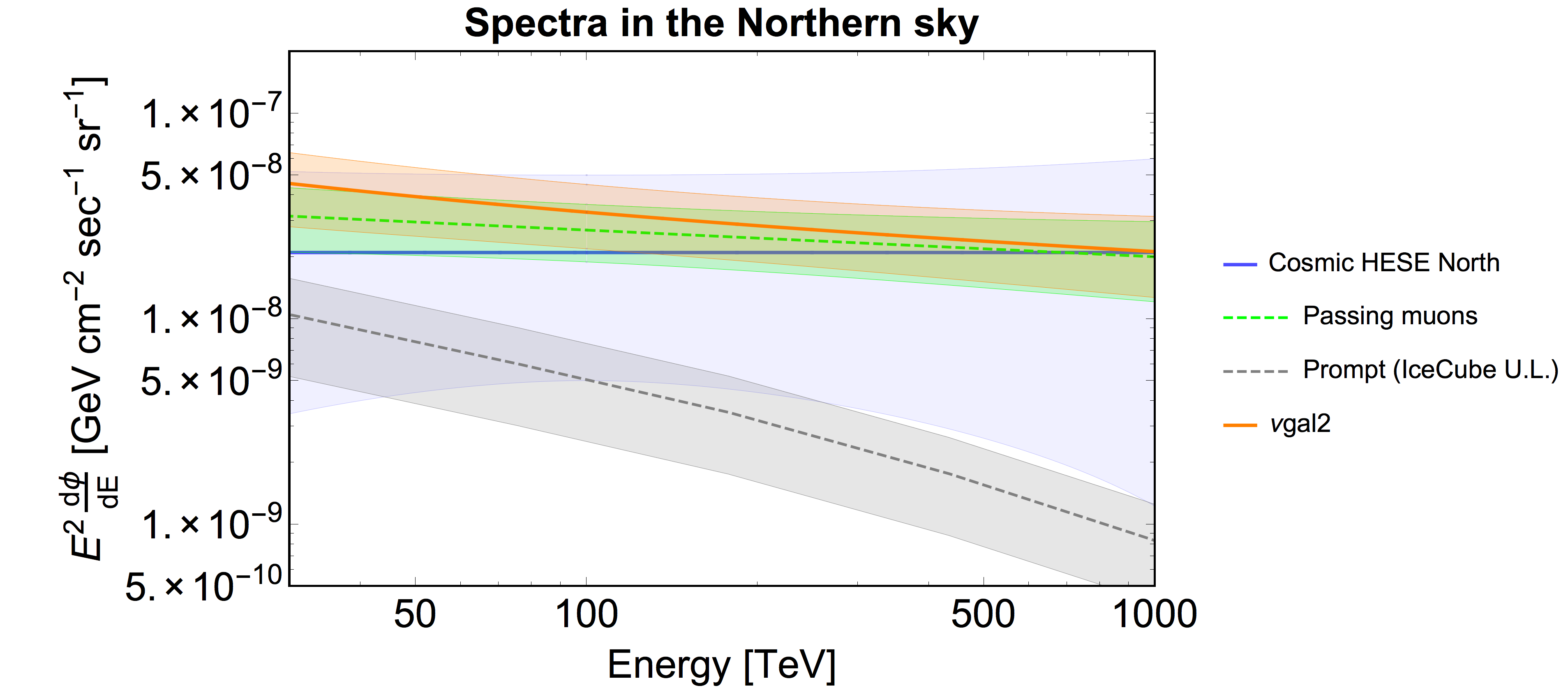}
\includegraphics[width=0.495\linewidth]{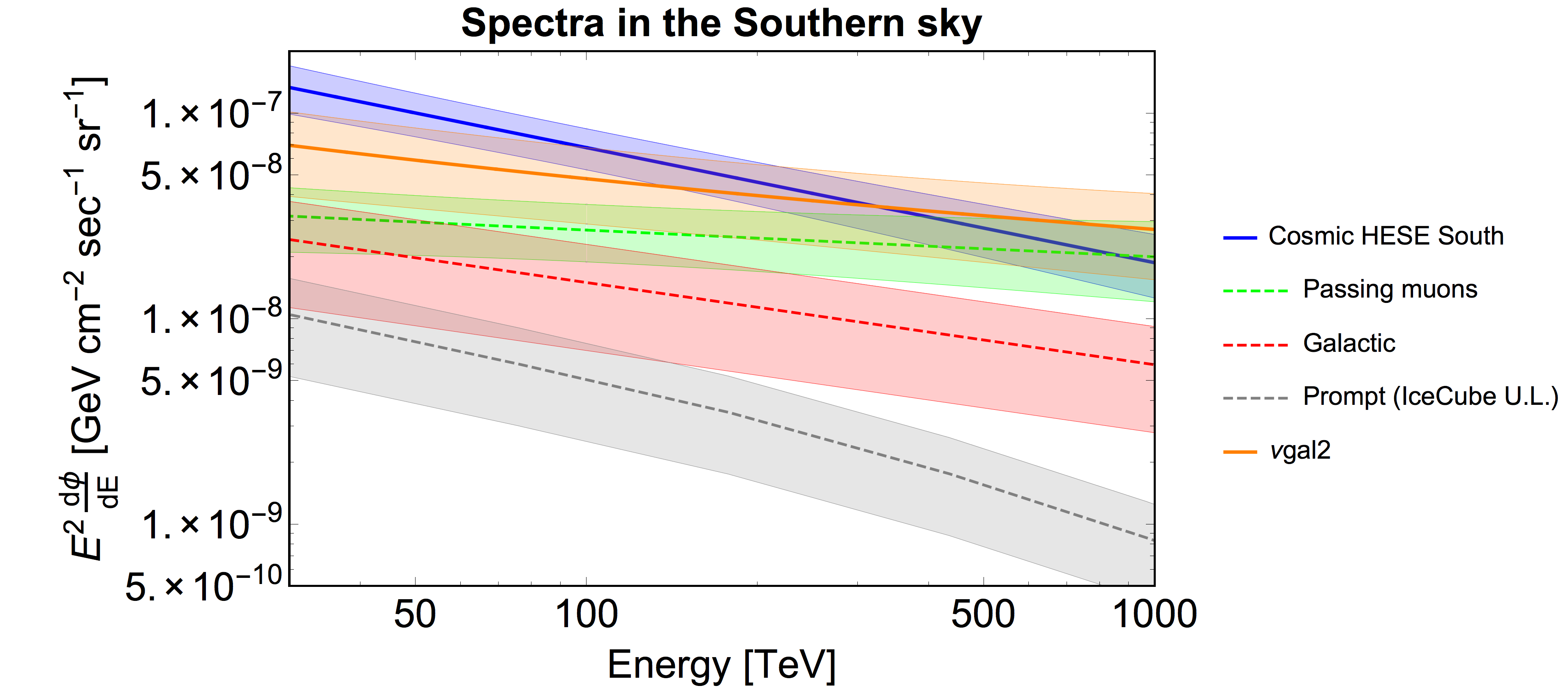}
\caption{\textit{The different components of the high-energy neutrino flux in the $\nu${\sf gal2} model.  
For comparison, the single power-law fits obtained by the IceCube collaboration are also given (light blue).}}
\label{spectra}
\end{figure*}
%%%%%%%%%%%%%%%%%%%%%%%%%%%%%%%%%%%%%%%%%%%%%%%%%%%%%%%%%%%%%%%%%%%%%%%

The expected spectra for the different components are reported in Fig.~\ref{spectra}, separated for the Northern and for the Southern hemisphere. The bands are due to the uncertainties on the normalization and on the spectral index.
More precisely, the upper limit of the bands is obtained with the following formula for the increment, 
\begin{equation}
\sqrt{[\phi(F,\alpha)-\phi(F^+,\alpha)]^2+[\phi(F,\alpha)-\phi(F,\alpha^+)]^2}
\end{equation}
where $\phi$ is the flux at best fit, $F^+=F+\Delta F$ and $\alpha^+=\alpha+\Delta \alpha$ if $E_\nu < \rm 100 \ TeV$ or $\alpha^+=\alpha-\Delta \alpha$ if $E_\nu > \rm 100 \  TeV$. 
The lower limit of the bands is obtained with the same procedure.

In Table \ref{tabev} there are 8.8 extragalactic events expected from the Northern hemisphere and 16.1 extragalactic events from the Southern hemisphere. Since the solid angle covered by the region with $|b| \geq 15^\circ$ is about 3/4 of $4\pi$, under the hypothesis of isotropic flux we expect to see $\sim$ 6.5 events from North and $\sim$ 12 events from South, coming outside the Galactic plane. Observing the map in Fig.\ref{hesegal} we found 12.5 events in Southern sky and 5.5 events in Northern sky (1 event is at declination $\delta=0^\circ$), in good agreement with the expectations.

%%%%%%%%%%%%%%%%%%%%%%%%%%%%%%%%%%%%% 
\subsection{Constraints from the ANTARES telescope}
%%%%%%%%%%%%%%%%%%%%%%%%%%%%%%%%%%%%%
A Galactic neutrino component is likely not isotopically distributed over the $2\pi$ sr of the Southern sky.
In the following, we will assume that Galactic neutrinos are produced in a rectangular region symmetrically extended with respect to the Galactic center up to Galactic latitude $|b|^*$ and longitude $|\ell|^*$. The solid angle covered by this region corresponds to
\begin{equation}\label{eq:dw}
\Delta \Omega^*= 4 \sin{|b|^*} \cdot |\ell|^* \ .
\end{equation}
Under this assumption, the normalization factor $ F_{0,\mbox{\tiny g}}$ given in (\ref{eq:PV2}) is null outside this rectangular region and inside becomes 
\begin{equation}\label{eq:f0star}
F^*_{0,\mbox{\tiny g}} = F_{0,\mbox{\tiny g}} \cdot {2\pi \over \Delta \Omega^* }
\end{equation}

The ANTARES neutrino telescope \cite{fusco} searched for an excess of events with respect to the background of atmospheric neutrinos in a region with $|b|^*=3^\circ$ and $|\ell|^*=40^\circ$, corresponding to $\Delta \Omega^* = 0.145$ sr.
From the null observation, and assuming a spectral index $\alpha_g=2.4$, a 90\% C.L. upper limit on the normalization for one-flavor neutrino of $\Phi_{Antares}^{1f}=2.0\times 10^{-17}$ $({\rm GeV \ cm^{2} \ s \ sr })^{-1}$ was derived.
Assuming a 1:1:1 flavor ratio, this value must multiplied by a factor of three to give the 90\% c.l. upper limit on all neutrino flavors, $\Phi_{Antares}^{3f}$. 
When compared with eq. \ref{eq:f0star} for $\Delta \Omega^* =0.145$ sr, the predicted flux 
$ F^*_{0,\mbox{\tiny g}} =6.5 \times 10^{-17}$ $({\rm GeV \ cm^{2} \ s \ sr })^{-1}$, is above $\Phi_{Antares}^{3f}$.
Thus, at 90\% C.L. it is excluded that the region originating Galactic neutrinos has dimensions smaller than $|b|^*<3^\circ$ and  $|\ell|^*<40^\circ$.
Larger production regions are still compatible with the ANTARES limits.

In Tab.\ref{GCx} we considered three rectangular regions with latitude $|b|^*=4^\circ$ and different values of Galactic longitude $|\ell|^*$. Here, we evaluated the fraction of the Galactic neutrino flux seen in different intervals of declination $\delta$.  For the IceCube geographical location in the South Pole, remember that the elevation corresponds to the declination: $\delta>0\ (<0)$ correspond to upgoing (downgoing) events.    

We can notice that the minimal hypothesis in which the Galactic flux is only seen from the Southern hemisphere ( $\delta<0$) is valid for a $|\ell|^* \leq 30^\circ$.
IceCube can observe this sky region only using the contained events (HESE) that have a poor angular resolution.
If the Galactic flux extends at larger longitudes, IceCube has the possibility to detect the signal also using upgoing muons. For instance, if $|\ell|^* = 70^\circ$, about 1/4 of the flux would be originating in the Northern hemisphere. 
However, since the flux measured by IceCube from the Northern hemisphere, both with passing muons and HESE, seems to have spectral index very close to $\alpha=2$, it is plausible that the Galactic flux has a longitudinal extension not exceeding $50^\circ$. 

In addition to tracks, ANTARES is studying the Galactic center region also using cascade events, mainly induced by CC $\nu_e$ and neutral current interactions \cite{grego_nu16}. The angular resolution of upgoing showering events in ANTARES is about $3^\circ-4^\circ$, thus comparable with the considered minimal extension of the rectangular region in Galactic longitude.
This will allow to test in a short timescale our scenario of Galactic neutrino production. 
Finally, the incoming experiment KM3NeT \cite{km3} has the chance to give the final answer to the existence of such Galactic neutrino flux, since it can observe most of the Galactic plane using passing muons and cascade events with an even better angular resolution than ANTARES. 

%%%%%%%%%%%%%%%%%%%%%%%%%%%%%%%%%%%%% 
\subsection{Remarks and conclusions}
%%%%%%%%%%%%%%%%%%%%%%%%%%%%%%%%%%%%%
The agreement of the $\nu${\sf gal2} with the observations is good, both for the Northern and for the Southern hemisphere. Indeed the total number of HESE observed by IceCube is compatible with our predictions (Table ~\ref{tabev}) within 1$\sigma$. 
Moreover we can notice that the extragalactic contribution is the larger component of the HESE events seen from the Southern sky, but it amount to about half of the total number of events. 
The remaining part is attributed to conventional neutrinos, atmospheric muons, prompt neutrinos and Galactic neutrinos.
Also assuming the maximum allowed flux, prompt neutrinos cannot explain the difference between the spectra derived from the passing muons and HESE in the Southern sky.

In Fig.~\ref{spectra} the fluxes predicted by our model are compared with the fluxes observed by IceCube in Northern and Southern hemisphere. A good agreement can be noticed below $\approx$ 100 TeV; it means that not only the $E^{-2.7}$ of \cite{apj} but also an $E^{-2.4}$ can adequately explain the low energy data. We conclude that the detailed shape of the Galactic spectrum is not crucial for the argument, 
whereas the presence of a significant Galactic component (i.e., of a normalization sufficiently large) is instead 
necessary to explain the North-South asymmetry and to soften the spectrum of the Southern hemisphere at low energy.

\begin{figure}[t]
\centering
\includegraphics[width=0.8\linewidth]{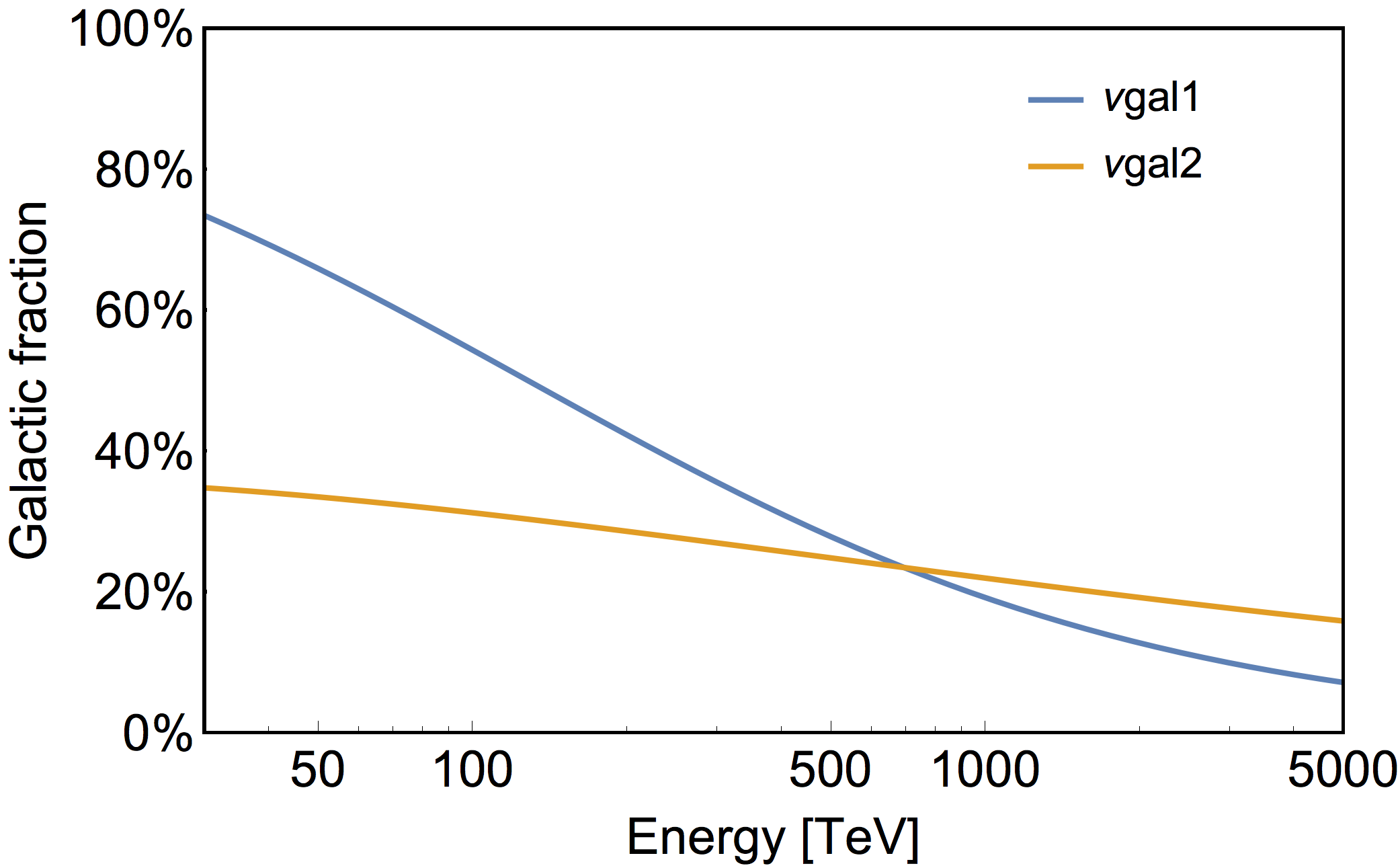}
\caption{\textit{Fraction of the neutrino emission in the Southern sky \andrea{(after removing the conventional background and the muons)} due to a Galactic
emission, as function of neutrino energy.}}
\label{fig:ggg}
\end{figure}

Finally, Fig.\ref{fig:ggg} shows the contribution of the Galactic component $\phi_{\mbox{\tiny g}}$ to the signal ($\phi{\mbox{\tiny g}}+\phi{\mbox{\tiny eg}}+\phi{\mbox{\tiny prompt}}$ or only $\phi{\mbox{\tiny g}}+\phi{\mbox{\tiny eg}}$ in $\nu${\sf gal1}) in the Southern sky; we can notice that the prediction of $\nu${\sf gal2} model (this paper) differs from $\nu${\sf gal1} model (\cite{apj}) by about a factor 2 at low energy, due to the fact that here a prompt neutrino component is taken into account and the spectrum of Galactic neutrinos is harder. 

%%%%%%%%%%%%%%%%%%%%%%%%%%%%%%%%%%%%%%%%%%%%%%%%%%%%%%%%%%%%%%%%%%
%%%%%%%%%%%%%%%%%%%%%%%%%%%%%%%%%%%%%%%%%%%%%%%%%%%%%%%%%%%%%%%%%%

%%%%%%%%%%%%%%%%%%%%%%%%%%%%%%%%%%%%%%%%%%%%%%%%%%%%%%%%%%%%%%%%%%%%%%%%% 
\section{A new isotropic component at low energies?\label{cacaton}}
%%%%%%%%%%%%%%%%%%%%%%%%%%%%%%%%%%%%%%%%%%%%%%%%%%%%%%%%%%%%%%%%%%%%%%%%% 

In this section, we discuss an alternative hypothesis that reconciles the observed spectra of the passing muons and HESE data. It assumes that there are two isotropic components of the extragalactic neutrinos, one with a hard spectral index needed to account for the high-energy part of the spectrum  and the second with a softer index, to explain the excess of HESE observed in the Southern sky.
To the best of our knowledge, this hypothesis was proposed in \cite{aron} and compared with the HESE data in \cite{Vincent:2016nut}, arguing that the existing HESE  data do not constrain it significantly.  Furthermore,  the passing muons \cite{Passing} have energies larger than 190 TeV, so they do not probe the crucial region of energies.

%However, a new and valuable information was recently obtained.  This information derives from the dataset of muon neutrinos with energies from few 100 of GeV till few 100 of TeV, that have been collected {\em from the Northern sky}  and have been just analyzed by IceCube~\cite{Passing}.This is the same analysis recalled in Sect~\ref{paralla}, aimed to probe prompt neutrinos with a known spectrum. The principle of the analysis is just that the prompt spectrum (approximatively a power-law with spectral index close to  $\alpha=2.7$) is significantly different from the spectrum of the conventional component  (close to a power-law   with $\alpha=3.7$ instead);  the former component should be visible over the latter one at  sufficiently high energies,  depending upon the--largely unknown--normalization. This analysis allowed IceCube collaboration to obtain a tight limit on  prompt neutrinos.}

To discuss the hypothesis, we use the valuable information recently added in the IceCube paper of passing muons ~\cite{Passing}, using muon neutrino events from few 100 of GeV until few 100 of TeV, collected {\em from the Northern sky}.
They considered that the spectrum of prompt neutrinos has spectral index 
$\alpha\simeq 2.7$, significantly different from that of the conventional component ($\alpha=3.7$).
The former component should overcome the latter at sufficiently high energies, depending upon the--largely unknown--normalization. 
This analysis allowed IceCube collaboration to obtain the upper limit reported in Sect. \ref{paralla}.
This constraint for events coming from the Northern hemisphere does apply directly to any new isotropic population with $E^{-2.7}$ spectrum, since (by assumption) muon neutrinos should be present in both hemispheres.  

In a very general way, we can thus consider and compare two assumptions: 
The first one is  the hypothesis of a new Galactic component, confined in the Southern sky; the second one is the hypothesis of a new isotropic and {\em soft} component, adding up to the atmospheric prompt neutrinos.
Referring to our Table~\ref{tabev}, it will be equivalent to replace the two rows ``Galactic'' and ``Prompt'' with one ``extragalactic soft''.
The spectral index of this soft component is unknown, but by assumption not largely far from the interval $\alpha \approx 2.4 \div 2.7$. 

The IceCube upper limit (obtained with the passing muon sample below 100 TeV) imposes that the normalization of the flux from the Northern sky is at maximum $F_{0,\mbox{\tiny prompt}}$, reported in Eq. \ref{eq:PV2}. This yields the number of HESE in the North reported in Table~\ref{tabev}, i.e. about 1.5 events.
As this extragalactic component is isotropic, at most $2\pm 1$ HESE are thus expected in the South. This expectation does not change within error if the spectral index ranges in the mentioned interval.
In conclusion, this second hypothesis is constrained by IceCube observation itself to produce at most $ 2$ HESE in the Southern sky, instead of the 6+2 events foreseen by the presence of a Galactic component. 
%Also the distribution of events as a function of the Galactic latitude %(Sect. \ref{norma}) seems better in agreement a

Thus, the IceCube spectral anomaly seems better explained by the presence of a component observable only in the Southern hemisphere, because this is not constrained from the considerations on prompt neutrinos. 
Further dedicated analyses of muon neutrinos (HESE and passing) and of showering events at low energy, assuming that the spectral index is in the range $\alpha=2.4-2.7$  will allow to constrain more precisely this alternative assumption.
Evaluable additional information could arise also from telescopes located in the Northern hemisphere.
Here, in view of the above argument, we assume that the new isotropic component is negligible or absent.

%\del{Evidently, this result has no bearing on the existence of Galactic neutrinos from the Southern sky. However, it does apply directly to a new isotropic population with  $E^{-2.7}$ spectrum, since (by assumption) the muon neutrinos should be present in both hemispheres.  In view of this consideration, and of the very tight bound of IceCube~\cite{Passing}, we conclude that it is not plausible that a new isotropic neutrino component  at low energies, distributed as $E^{-2.7}$, can reconcile the various observations of IceCube. 

%More generally, we can consider and compare two assumptions: The first one is  the hypothesis of a new Galactic component,  confined in the Southern sky; the second one is the hypothesis of a new isotropic component, with a softer energy spectrum, described by a power-law with spectral index $2.4\le \alpha\le 2.7$ smaller than the value discussed just above. 

%Although both of them can reconcile the IceCube spectral anomaly, in the second case it is to be expected that the difference of the new component and the conventional component (muon neutrinos and muons) will remain quite large and observable by the same sample analyzed in \cite{Passing}, i.e., the up-going muons with energies from few 100 of GeV till few 100 of~TeV. 
%This constraint does not apply to the first hypothesis, that is the basis of the PV1 and of the PV2 models. }

%%%%%%%%%%%%%%%%%%%%%%%%%%%%%%%%%%%%%%%%%%%%%%%%%%%%%%%%%%%%%%%%%%%%%%%%% 
\section{Summary and discussion\label{conc}}
%%%%%%%%%%%%%%%%%%%%%%%%%%%%%%%%%%%%%%%%%%%%%%%%%%%%%%%%%%%%%%%%%%%%%%%%% 

In this paper we have discussed a general model to explain the North-South asymmetry whose indications are growing up from IceCube data. 
We have analyzed the so called $\nu${\sf gal2}, that generalizes the model with extragalactic and Galactic components for cosmic neutrinos, as motivated in Sects.~\ref{high} and \ref{pv2sum}.
We have considered:
current theoretical expectations;
the presence of the prompt neutrino component; 
the effect of the spatial distribution of the Galactic neutrinos; 
the impact of the new observations of the passing muons above few 100 TeV and from the Northern sky, which we assume to derive from an   extragalactic component; 
the constraints arising from the ANTARES neutrino telescope from neutrinos from the Southern hemisphere.
These modifications have been justified by various theoretical and observational reasons.

% furthermore,  we aimed to  maximize the differences with PV1, in order to test the stability of the conclusions reached in \cite{apj}, {with the purpose to explain the spectral anomaly observed by IceCube.

Different results arise from the discussion of the model.
%The main results of the discussion of Sect.~\ref{sec:unk} 
%are two,

1) A Galactic component, assumed with energy distribution $E^{-2.4}$ (as motivated in Sect.~\ref{slopa}) and with a smaller normalization coefficient than assumed in the $\nu${\sf gal1} (as discussed in Sect.~\ref{sommariomod}) is also compatible with the existing observations. 
It explains the distribution of the HESE from the Southern sky, receiving both Galactic and extragalactic contributions. 
In the model, HESE from the North are induced only by extragalactic neutrinos, whose flux is measured from the passing muon sample.

%The main  difference with the PV1 model is that the HESE events observed from the Southern sky should have a different spectral index: 

2) Assuming a spectral index of 2.4 instead of 2.7 as in the $\nu${\sf gal1} case the fraction of Galactic neutrino events at the lowest energies diminishes, whereas we have a relatively large amount of Galactic neutrinos at higher energies, as emphasized in Fig.~\ref{fig:ggg}
\footnote{Note incidentally that if the knee of the cosmic ray spectrum (observed at the Earth) is due to a feature of the accelerators, we would expect to have a corresponding feature in the Galactic neutrinos at energies of 
$\epsilon_\nu \times E_{\mbox{\tiny knee}} \approx 1/20\times \mbox{3 PeV}=  150$ TeV: this makes it evident how important is a detailed study of the Galactic neutrinos emission.}
. This increases the probability that an excess of events with respect atmospheric neutrinos can be seen by the ANTARES detector.

3) The Milky Way is mostly visible from the Southern sky (Table \ref{GCx}). This means that the Galactic component could be measured in IceCube using HESE only, while is accessible as upgoing muons and cascades in neutrino telescopes located in the Northern hemisphere.

4) The prompt neutrino flux cannot produce a North-South asymmetry (Sect. \ref{cacaton}), but it could modify the shape of the spectrum, making it softer at low energies.  
However, after the recent bounds obtained by IceCube, its presence is too small to modify significantly the HESE spectrum, thereby accounting for the discrepancy between the HESE South data and the passing muon data. 
This conclusion becomes even stronger if the most recent calculations of prompt neutrinos~\cite{p1,p2,p3} are assumed.

A crucial dataset used to test this model--in any of its variant, $\nu${\sf gal1}, $\nu${\sf gal2}, or others--is the IceCube HESE from the Southern sky and below few 100 of TeV. Moreover, a common feature of all the $\nu{\sf gal}$ models is that most of the cosmic neutrino events at high Galactic latitude are of extragalactic origin. 
Therefore, the subset of the HESE North and South that come from a Galactic latitude $|b| \ge 10^\circ-15^\circ$ (the angular resolution of the showers) should be consistent between them.
%and should be  explained by the same neutrino spectrum that accounts for the passing muon dataset. 
It is important that, in the future, the IceCube collaboration perform similar tests.
%\andyb{It could be interesting a spectral analysis, performed by the IceCube collaboration, of the HESE from Southern hemisphere coming from a Galactic latitude $|b| \geq 15^\circ$, to test the agreement with the HESE from Northern hemisphere. Indeed the agreement between these two datasets would be another important hint to support the existence of the Galactic component of high energy neutrinos.}
% (see Sect.~\ref{high} and in particular the subsection~\ref{pipr} for a %discussion of the high energy part).  
Referring to Table~\ref{tabev}, a potential problem arises if some background process, caused or connected to atmospheric muons, is not fully under control as currently expected:  in this case, the reliability of the conclusions should be reassessed.
In any case, the existing statistics is still insufficient to be certain  of a significant Galactic neutrino emission, and a firmer conclusion can be obtained with more HESE data (still on IceCube disks) and with data collected from the direction of the Galactic plane by Mediterranean neutrino telescopes (ANTARES and KM3NeT in the near future).

%But even assuming a strict correctness of the analyses of IceCube collaboration, as we did in the present work, it should be said that the existing statistics is still insufficient to be certain  of a significant Galactic neutrino emission. More HESE data and an improved analysis of the atmospheric muons are needed to derive a reliable conclusion.

%However, as we have argued in \cite{apj} and in the present work, the hypothesis that the Galactic neutrino emission is significant and potentially observable by IceCube enjoys the following properties,
%\begin{enumerate} 
%\item it is consistent with current theoretical expectations;
%\item it is supported/suggested by various pieces of data collected by IceCube, in particular by the comparison of the HESE events  from the Southern sky and the passing muon events from the Northern sky; 
%\item it has some advantages in comparison with other hypotheses  (see sect.~\ref{cacaton})
%\item and finally and quite importantly, it is rather stable under several variations of the parameters of the model, as the ones considered in the present work.
%\end{enumerate}

\begin{acknowledgements}
We thank C.~Evoli, F.~Halzen, P.~Sapienza, F.~Tavecchio and W.~Winter for precious discussions.
\end{acknowledgements}

%\section{}
%\subsection{}

\end{document}